\documentclass[
reprint,
superscriptaddress,
amsmath,amssymb,
aps,
pra
]{revtex4-2}

\usepackage{graphicx}
\usepackage{dcolumn}
\usepackage{bm}
\usepackage[colorlinks=true, citecolor=blue, linkcolor=blue]{hyperref}
\usepackage{color}
\usepackage[capitalize]{cleveref} 
\usepackage{siunitx}
\newcommand{\ud}[1]{{#1^{\dagger}}}
\newcommand{\av}[1]{{\langle #1 \rangle}}
\usepackage{physics}
\usepackage[T1]{fontenc}

\begin{document}

\title{Selective filtering of multi-photon events from a single-photon emitter}

\author{Friedrich Sbresny}
 \email{Corresponding author: friedrich.sbresny@tum.de}
 \affiliation{Walter Schottky Institut, TUM School of Computation, Information and Technology, and MCQST, Technische Universität München, 85748 Garching, Germany}
 \author{Carolin Calcagno}
 \affiliation{Walter Schottky Institut, TUM School of Computation, Information and Technology, and MCQST, Technische Universität München, 85748 Garching, Germany}
\author{Sang Kyu Kim}
 \affiliation{Walter Schottky Institut, TUM School of Computation, Information and Technology, and MCQST, Technische Universität München, 85748 Garching, Germany}
\author{Katarina Boos}
 \affiliation{Walter Schottky Institut, TUM School of Computation, Information and Technology, and MCQST, Technische Universität München, 85748 Garching, Germany}
\author{William Rauhaus}
 \affiliation{Walter Schottky Institut, TUM School of Computation, Information and Technology, and MCQST, Technische Universität München, 85748 Garching, Germany}
\author{Frederik Bopp}
 \affiliation{Walter Schottky Institut, TUM School of Natural Sciences, and MCQST, Technische Universität München, 85748 Garching, Germany}
\author{Hubert Riedl}
 \affiliation{Walter Schottky Institut, TUM School of Natural Sciences, and MCQST, Technische Universität München, 85748 Garching, Germany}
\author{Jonathan J. Finley}
 \affiliation{Walter Schottky Institut, TUM School of Natural Sciences, and MCQST, Technische Universität München, 85748 Garching, Germany}
\author{Eduardo Zubizarreta Casalengua}
 \affiliation{Walter Schottky Institut, TUM School of Computation, Information and Technology, and MCQST, Technische Universität München, 85748 Garching, Germany}
\author{Kai Müller}
 \affiliation{Walter Schottky Institut, TUM School of Computation, Information and Technology, and MCQST, Technische Universität München, 85748 Garching, Germany}

\date{\today}

\begin{abstract}
Single-photon purity is one of the most important key metrics of many quantum states of light.
For applications in photonic quantum technologies, e.g. quantum communication and linear optical quantum computing, a minimization of the multi-photon error rate is required because of its error-introducing nature.
Ultimately, the purity of state-of-the-art single-photon sources was found to be limited by spontaneous emission and subsequent reexcitation during the interaction with the driving field.
Here, we demonstrate that even this fundamental limit to the single-photon purity can be overcome due to the distinct spectro-temporal properties of the individual photons forming multi-photon errors.
For driving pulses shorter than the emitter lifetime, we find that photons emitted during the pulse exhibit a significantly broader spectral shape than the emitter's natural linewidth.
Thus, we can selectively filter out the majority of this instantaneously emitted photon by employing narrowband spectral filters which reduces the measured degree of second-order coherence at zero time delay by almost one order of magnitude.
This enables a significant suppression of the multi-photon error rate without detrimental effects on the desired single-photon emission.
\end{abstract}
\maketitle
Single photons are the most fundamental quantum resource for applications in photonic quantum technologies.
They are a key requirement for quantum secure communication \cite{bennett2014}, linear optical quantum computing \cite{knill2001, kok2007} and boson sampling \cite{aaronson2011, zhong2020}.
In addition, single photons form the basis for more exotic quantum states of lights, such as entangled photon pairs \cite{kocher1967,stevenson2006} and cluster states \cite{lindner2009,schwartz2016,lee2019, thomas2022}.
For such non-classical states of light, the single-photon purity remains a crucial property to minimize errors in the implementation of protocols e.g. quantum repeater schemes for quantum communication \cite{briegel1998,azuma2015,azuma2023} and measurement-based quantum computing \cite{raussendorf2003,briegel2009}.
Extensive research has been carried out to optimize the single-photon generation and minimize the multi-photon error rate with a strong focus on advancing quantum emitters and excitation schemes \cite{higginbottom2016,somaschi2016,senellart2017,trivedi2020,uppu2020, tomm2021, rota2024, rickert2025, ding2025}.
It was found that for any level system driven by a pulsed electromagnetic field, the best obtainable single-photon purity is ultimately limited by spontaneous emission and subsequent reexcitation during the system-pulse interaction, where the first emitted photon is responsible for introducing the unwanted two-photon errors \cite{fischer2018a, hanschke2018}.
In this context, one fundamental concept of quantum mechanics has been paid little attention to, namely the measurement apparatus used to collect and detect the photons itself \cite{gonzalez-ruiz2025}. 
Since the measurement of a quantum mechanical state gives only one definite result, any component in the apparatus has the power to affect the outcome and thus requires a closer inspection. 
To close the gap between excitation of the quantum emitter and detection of the single photons, here we provide an in-depth study of the first emitted photon, finding that its spectral shape is ultimately determined by a combination of the driving pulse duration and the radiative lifetime of the emitter.
Hence, for relatively short pulses, in this context meaning shorter than the radiative lifetime of the emitter, the first photon is instantaneously emitted and therefore exhibits a broad spectral shape.
This fact allows selective spectral filtering of most of the instantaneous photon, thus reducing the multi-photon error rate by almost one order of magnitude without significantly attenuating the desired single-photon emission.

We illustrate this concept with a theoretical model for the two-level system, the most intuitive quantum system for single-photon generation.
In addition, we demonstrate this effect experimentally for a quantum four-level system with a cascaded decay in a semiconductor quantum dot, obtaining a minimum degree of second-order coherence at zero time delay of $g^{(2)}\left[0\right] = \SI[separate-uncertainty=true]{7.5(20)e-5}{}$.
To the best of our knowledge, this value is in line with the best single-photon purity obtained so far for such a system \cite{schweickert2018a}.
Both theory and experiment together give a coherent picture of the temporal and spectral nature and the origin of the instantaneous photon and provide an easily accessible experimental method to mitigate its negative impact on the multi-photon error rate by precise engineering of the spectral filter.
\section*{Results}
\subsection*{Spectro-temporal properties of multi-photon events}
The most fundamental single-photon source is based on a quantum two-level system.
Upon excitation with a resonant light field, its population undergoes coherent oscillations between its ground and excited state.
\begin{figure}
    \centering
    \includegraphics{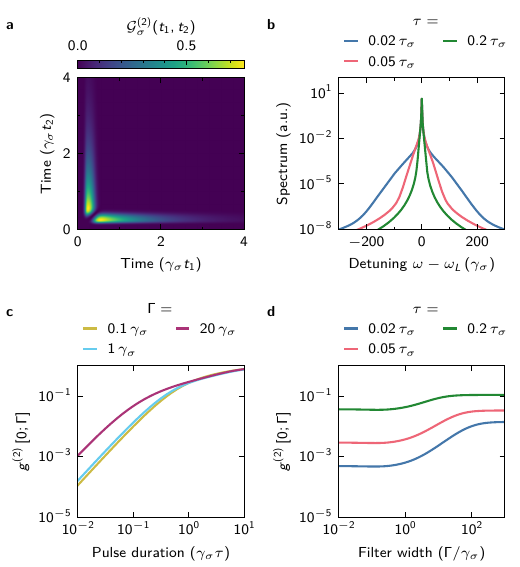}
    \caption{\textbf{Temporal and spectral shape of two-photon errors.} \textbf{a}, Calculated unnormalized two-time second-order correlation function of the two-level system upon excitation with a pulse of area $\pi$. Two-photon coincidences never occur at the same times but only in the specific time areas formed by the pulse width and emitter lifetime. \textbf{b}, Calculated spectral profiles of the emitter on a logarithmic scale. For shorter pulse durations, clear shoulders emerge from the main emission line. \textbf{c}, Normalized second-order coherence $g^{(2)}\left[ 0; \Gamma\right]$ of the two-level system emission in dependence of the pulse duration for three different filter widths $\Gamma= 0.1 \gamma_\sigma, 1 \gamma_\sigma, 20 \gamma_\sigma$. With decreasing pulse duration the $g^{(2)}\left[ 0; \Gamma\right]$ decreases due to a reduced reexcitation probability. For narrower filters, a significantly lower $g^{(2)}\left[ 0; \Gamma\right]$ is obtainable. \textbf{d}, Normalized second-order coherence $g^{(2)}\left[ 0; \Gamma\right]$ of the two-level system emission in dependence of the filter width for three different driving pulse durations $\tau = 0.02 \tau_0, 0.05 \tau_0, 0.2 \tau_0$. Decreasing the filter width significantly improves the single-photon purity for pulse lengths shorter than the emitter lifetime. }
    \label{fig:theory}
\end{figure}
By carefully choosing a pulse duration corresponding to only half an oscillation period, one can deterministically prepare the system in the excited state, from where it decays emitting only one photon at a time.
Multi-photon errors are introduced when the excited state decays prematurely and instantaneously emits one photon during the pulse duration, leaving the possibility for the remaining pulse area to reexcite the system resulting in the subsequent emission of a second photon \cite{fischer2018a}.

In the temporal domain, the relationship between the instantaneously emitted photon and the desired single photon (sometimes unprecisely coined reexcitation photon because of its temporal occurence after the first photon) can be investigated in the unnormalized two-time second-order correlation function $\mathcal{G}_{\sigma}^{(2)}(t_1, t_2)$ of the two-level system under pulsed excitation.
This is displayed in \cref{fig:theory}a, where the two time axes are normalized to the emitter decay rate $\gamma_\sigma= 1/ \tau_\sigma$, which is the inverse of the emitter decay time. 
Upon driving the system with a Gaussian pulse centered at time $t= 0.2 \tau_{\sigma}$, the first photon is detected at either detection time $t_1$ or $t_2$.
For an excitation pulse of duration $\tau= 0.05 \tau_{\sigma}$, this occurs within the short time frame of the excitation pulse, determining the narrow shape of the coincidences along one axis.
The second photon is detected at the other detection time, leading to the observed symmetric shape with the correlation intensity modulation along the second axis corresponding to the characteristic monoexponential decay of the radiative lifetime.
This clearly illustrates that the two-photon events do not occur at the exact same time, but in contrast arise from subsequent excitations and emissions of the emitter during one driving cycle.

Since the first photon is emitted during the system-pulse interaction, it is spectrally broad compared to the second, lifetime-limited photon.
This results in a distinct spectral shape of the total emission which is displayed in the calculated logarithmic spectra in \cref{fig:theory}b for three different excitation pulse durations of $0.02 \tau_{\sigma}$ (blue),  $0.05 \tau_{\sigma}$ (red) and $0.2 \tau_{\sigma}$ (green).
With decreasing excitation pulse length, broad shoulders start to emerge in the spectra, which clearly relates their origin to the temporal width of the pulse and thus of the instantaneous photon. 
While the spectral shoulders absolute emission intensity compared to the main lifetime-limited peak at zero detuning $\omega-\omega_L=0$ is almost negligible, they are the main contribution to the multi-photon error rate.

\begin{figure*}
    \centering
    \includegraphics{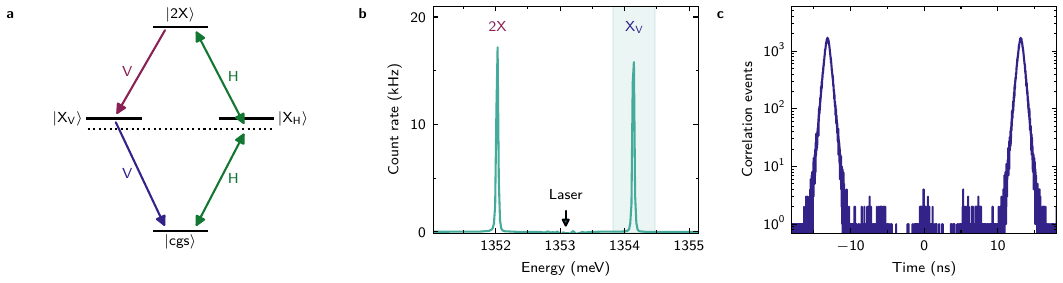}
    \caption{\textbf{Cascaded four-level system as a single-photon emitter.} \textbf{a}, Level scheme of the four-level system comprised of the biexciton state ($\ket{\mathrm{2X}}$), two fine structure-split exciton states ($\ket{\mathrm{X}_\mathrm{V}}$ and $\ket{\mathrm{X}_\mathrm{H}}$) and the crystal ground state ($\ket{\mathrm{cgs}}$). The system is driven resonantly via the two-photon resonance of the horizontally polarized branch (two green arrows). The cascaded emission of the biexciton (wine arrow) and exciton (indigo arrow) is collected in vertical polarization. \textbf{b}, Measured emission spectrum of the vertically polarized cascaded decay under $\pi$-pulse excitation shows both the biexciton and exciton emission lines. The laser resonant on the two-photon transition is not visible due to cross-polarized suppression. \textbf{c} Second-order intensity autocorrelation measurement of the exciton emission filtered with a bandwidth of $\SI{159}{\giga \hertz}$ (shaded region in \textbf{b}) shows a clear antibunching at zero time delay. Here, a \SI{5}{\pico \second} intensity full width at half maximum long pulse was used to excite the biexciton.}
    \label{fig:basics}
\end{figure*}
Due to their broad frequency profile of multiples of the emitter linewidth, we can spectrally filter out the shoulders of the spectrum and consequently with them a significant amount of multi-photon events.
This is shown in \cref{fig:theory}c, where we plot the $g^{(2)}\left[0; \Gamma \right]$ in dependence of the excitation pulse length normalized to the emitter lifetime for three Lorentzian filters with bandwidths of $\Gamma=0.1 \gamma_\sigma$ (sand), $\Gamma=1 \gamma_{\sigma}$ (cyan) and $\Gamma=20 \gamma_\sigma$ (purple).
For excitation pulse durations longer than the radiative decay rate, we observe that the the single-photon purity is independent on the filter width. 
On the other hand, for pulse durations shorter than the radiative recombination time, the multi-photon errors are naturally suppressed by the reduction of the emission probability during the pulse.
The multi-photon error rate gets further reduced when reducing the filter width, with an improvement of the minimum obtainable $g^{(2)} \left[ 0; \Gamma \right]$  by almost one order of magnitude going from $\Gamma= 20 \gamma_\sigma$ (purple) to $\Gamma= 1 \gamma_\sigma$ (cyan).
Note that further narrowing the frequency filter to $\Gamma= 0.1 \gamma_\sigma$ (sand) does not significantly improve the $g^{(2)} \left[ 0; \Gamma \right]$, because the further reduction of the instantaneously emitted photon is payed for by a significant reduction of the desired single-photon emission.

The full potential of the particular spectro-temporal profile of the multi-photon emission becomes even clearer, when studying the $g^{(2)} \left[ 0; \Gamma \right]$ in dependence of the normalized filter width $\Gamma/ \gamma_\sigma$ for the excitation pulse durations $0.02 \tau_\sigma$ (blue), $0.05 \tau_\sigma$ (red) and $0.2 \tau_\sigma$ (green), as plotted in \cref{fig:theory}d.
For very large filter widths, the second-order correlation function at zero time delay plateaus at a constant value, because all multi-photon events contribute to $g^{(2)} \left[ 0; \Gamma \right]$.
When the spectral filter is narrowed down towards the emission rate of the emitter, the instantaneously emitted photon is partly filtered out resulting in a steep descent of $g^{(2)} \left[ 0; \Gamma \right]$.
For very narrow filters, much narrower than the emitter decay rate, we observe a second plateau where no further improvement in $g^{(2)} \left[ 0; \Gamma \right]$ is observed.
The absolute values of the $g^{(2)} \left[ 0; \Gamma \right]$ plateaus are lower for shorter pulse lengths, because the probability of photon emission during the pulse is already reduced.
Additionally, it can be observed that the decrease of multi-photon events on the logarithmic scale is more pronounced for shorter pulse durations, which indicates that more of the spectrally broader instantaneous photon can be filtered out.

The theoretical exploration of the parameter space yields a major finding: For excitation pulses shorter than the radiative lifetime, which is required for a high-purity single-photon source, the instantaneously emitted photon is spectrally broader than the natural emission.
This allows to selectively filter out mainly this component using filters wider than the emitter linewidth, significantly enhancing the single-photon purity.
\subsection*{High-purity single-photon emission from a cascaded decay}
The experimental realization of the improvement in  $g^{(2)} \left[ 0; \Gamma \right]$ after spectrally filtering is challenging under resonant excitation of a two-level system since laser and single photons have the same energy, which inhibits the high signal-to-noise ratio required for the observation of this effect (see Supplemental Material for more information).
To optimize this ratio, our system of choice for an experimental verification is the biexciton-exciton cascade in a semiconductor quantum dot since it unites high brightness and a very low multi-photon error rate with the advantages of resonant excitation and a spectrally separable emission \cite{brunner1994, stufler2006,hanschke2018,schweickert2018a}.
The level scheme of the two-path cascade in a semiconductor quantum dot, consisting of the crystal ground state ($\ket{\mathrm{cgs}}$), the biexciton state ($\ket{\mathrm{2X}}$) and two fine structure split exciton states ($\ket{\mathrm{X}_\mathrm{V}}$ and $\ket{\mathrm{X}_\mathrm{H}}$) is shown in \cref{fig:basics}a \cite{bayer2002}.
\begin{figure*}
    \centering
    \includegraphics{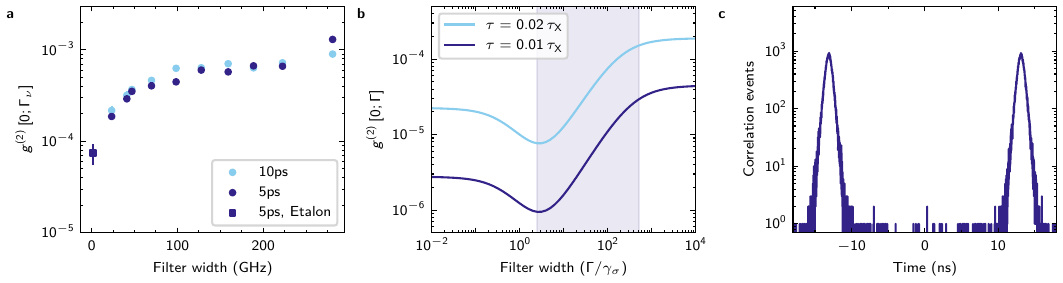}
    \caption{\textbf{Filter-dependent second-order coherence of the cascaded four-level system.} \textbf{a}, Measured degree of second-order coherence at zero time delay of the exciton transition in dependence of the filter width for $\SI{5}{\pico \second}$ and $\SI{10}{\pico \second}$ long pulses. Narrowing down the filter bandwidth significantly improves the measured $g^{(2)} \left[ 0; \Gamma_\nu \right]$. \textbf{b}, Simulated degree of second-order coherence at zero time delay of the exciton-cgs transition for pulse durations $\tau= 0.01 \tau_\mathrm{X}$ and $\tau= 0.02 \tau_\mathrm{X}$. The shaded region represents the in the experiment approximately accessible range of frequency filter widths. \textbf{c}, Second-order intensity autocorrelation measurement filtered by a narrow filter significantly improves the measured $g^{(2)}\left[ 0; \Gamma_\nu= \SI{1.4}{\giga \hertz} \right]= \SI[separate-uncertainty=true]{7.5(20)e-5}{}$ by selectively filtering out the instantaneous photon. }
    \label{fig:results}
\end{figure*}
Due to the biexciton binding energy, the biexciton-exciton transition is lower in energy than the exciton-cgs transition \cite{brunner1994}.
We efficiently excite the biexciton via a resonant two-photon absorption process using a horizontally polarized pulse of area $\pi$ (green arrows), which is spectrally red-detuned from the exciton transition. 
This allows us to deterministically prepare a biexciton, which then radiatively decays via the two paths.
Using polarization filtering, we look only at the vertically polarized transitions (wine and indigo arrows), where \cref{fig:basics}b shows a typical emission spectrum with the two distinct emission lines.
The laser pulse driving the system is centered between both transition lines but is strongly suppressed by cross-polarized filtering.
For an ideal four-level system both transitions should exhibit the same degree of second-order coherence.
However, for InGaAs quantum dots, which we study in this work, the tail of the broad phonon sideband of the exciton emission \cite{favero2003} leaks into the red-detuned biexciton emission peak.
This effect deteriorates the measurable second-order coherence of biexciton photons due to the intrinsic correlation of photons from the two transitions of the same cascade and dominates all other multi-photon errors. 

Thus, our transition of choice for the second-order correlation measurements is the $\ket{\mathrm{X}_\mathrm{V}}\rightarrow \ket{\mathrm{cgs}}$ transition, where the shaded region in the spectrum in \cref{fig:basics}b represents a spectral filter with bandwidth $\Gamma_\nu=\SI{159}{\giga \Hz}$,  which is wide compared to the Fourier-limited linewidth of $\SI{0.54}{\giga \hertz} $ extracted from the decay rate of the exciton. 
Filtering the emission of the exciton decay allows us to measure the second-order intensity autocorrelation using a Hanbury Brown and Twiss setup, with the result displayed in \cref{fig:basics}c.
In the logarithmic plot we clearly observe a strongly suppressed center peak at zero time delay corresponding to two-photon coincidences from a single pulse and the two neighboring peaks at $\pm \SI{13.1}{\nano \second}$ corresponding to coincidences of photons generated by two subsequent pulses. 
In addition, we observe the appearance of two additional peaks in between center and first neighboring peaks, which we identified as a measurement artifact originating from a bidirectional reflection inside our optical setup between the single photon detectors and the cryostat.
We determine the second-order coherence to be $g^{(2)}\left[ 0; \Gamma_\nu= \SI{159}{\giga \Hz}\right]= \SI[separate-uncertainty=true]{5.8(4)e-4}{}$.
The low value demonstrates the suppressed reexcitation of the four-level system, which arises from the low probability of both decays of the cascade happening within the pulse duration \cite{hanschke2018,schweickert2018a}.
\subsection*{Suppressing multi-photon errors by selective filtering}
Pairing our experimental approach with the theoretical predictions, we performed second-order coherence measurements for varying filter bandwidths.
The resulting curve is displayed in \cref{fig:results}a, where we plot the $g^{(2)} \left[ 0; \Gamma_\nu \right]$ in dependence of the filter bandwidth for two excitation pulses of  $\SI{5}{\pico \second}$ and $\SI{10}{\pico \second}$ intensity full width at half maximum.
For broad filters above \SI{100}{\giga \hertz} bandwidth, we observe that the $g^{(2)} \left[ 0; \Gamma_\nu \right]$ values saturate approximately at $\SI[separate-uncertainty=true]{6.5e-4}{}$, which gives a clear indication that the full spectrum of the instantaneous photon is within the filter bandwidth.
For narrower filters, $g^{(2)} \left[ 0; \Gamma_\nu \right]$ drastically decreases by almost one order of magnitude, with the filter always being broader than the radiative linewidth of the emitter.
This behavior is equivalent to the theoretical predictions.
The measured values of the shorter pulse durations are for all but two data points lower than for the longer pulse duration.
The exception at the plateau likely comes from a statistical fluctuation.
For the broadest filter of $\SI{281}{\giga \hertz} $ and the shorter pulse duration it originates from laser light of the tail of the excitation laser leaking into the emission, despite the spectral and polarization filters applied.

The theoretically calculated values of $g^{(2)} \left[ 0; \Gamma \right]$ for the four-level system as a function of the filter width can be seen in \cref{fig:results}b for the measured emitter lifetime and the studied pulse durations.
The shaded region represents the approximate area where the experimental measurements were realized.
We note that a quantitative comparison to the experiment is difficult, as the experimentally used filters have a complex spectral shape that depends on the filter width and thus deviates between all individual measurements.
Additionally, imperfect Gaussian pulse shapes, imperfect preparation fidelity and uncorrelated noise sources in the experiment may lead to further mismatch.
Nevertheless, in \cref{fig:results}b, we observe a similar trend as in the experiment with the characteristic plateau for large filter widths, and a reduction by approximately one order of magnitude for decreasing filter widths. 
Interestingly, for the four-level system under these excitation conditions, we observe a dip at a filter width of $\Gamma/\gamma_\sigma=2.5$ times the linewidth, while even narrower filters have a detrimental effect on the degree of second-order coherence.
This implies that narrowing down the filter below a certain width reduces the single-photon component faster than the instantaneous photon.

To maximally exploit the spectral multi-photon error suppression, we measure the degree of second-order coherence after adding an etalon with a bandwidth of $\SI{1.4}{\giga \hertz}$ to the narrowest grating based filter. This is represented as the square in \cref{fig:results}a, while the correlation measurement for this configuration is presented in \cref{fig:results}c. 
Here, we obtain a much lower $g^{(2)}\left[ 0; \Gamma_\nu= \SI{1.4}{\giga \hertz} \right]= \SI[separate-uncertainty=true]{7.5(20)e-5}{}$.
So far, such a low value was only achieved for a semiconductor quantum dot when also using short excitation pulses and a narrow spectral filter \cite{schweickert2018a}. 
Note that while the here used etalon of $\SI{1.4}{\giga \hertz}$ bandwidth is broader than our expected Fourier-limited linewidth of the emitter ($\SI{0.54}{\giga \Hz}$ extracted from the $\mathrm{X}_\mathrm{V}$ lifetime), the studied sample and setup are subject to residual periodic electric noise resulting in spectral diffusion, which gives rise to the observation of periodic blinking in this measurement. 
Since this effect can be mitigated by improving the wire shielding, DC sources with lower noise and electric lowpass filters, this measurement only poses an upper bound on the obtainable $g^{(2)} \left[ 0; \Gamma_\nu \right]$ in this configuration (see Supplemental Material for detailed information on the blinking).
\section*{Discussion}
In this work, we explored theoretically and experimentally the spectro-temporal properties of undesired multi-photon emission events of single-photon emitters.
This type of multi-photon events is characterized by the instantaneous emission of one photon during the driving pulse, with the subsequent excitation to a quantum superposition between ground and excited state with the remaining pulse area, allowing for the emission of a second photon.

For short excitation pulse lengths the instantaneous emission exhibits a distinct spectral shape, which is much broader than the natural emission linewidth but very weak in intensity.
This enables the selective spectral filtering of a significant portion of the instantaneous photon and with it the corresponding multi-photon events, reducing the measured degree of second-order coherence at zero time delay by almost one order of magnitude.
Since many practical real-world implementations require an optical filter, our findings allow to tailor the driving field duration and filter bandwidth to the optimum of both high single-photon transmission and low multi-photon error rate for the specific application.

Moreover, we believe that our results will have a major impact and require further investigations for single-photon emitters embedded in nanophotonic cavities \cite{somaschi2016,rota2024,ding2025}.
While cavities reduce the emitter lifetime via Purcell enhancement, their spectral transmission also affects the spectral and temporal shape of both the excitation pulse and the emission, thus significantly altering the second-order coherence measured from such devices.

We are confident that our findings can be used to further improve single-photon sources by not only focusing on the emitter, but the trio of the emitter, excitation technique and the measurement apparatus collecting the emission.
All of these aspects are in the end jointly responsible for the properties of the generated photon stream and we hope that this work will allow to establish more reproducible methods to benchmark single-photon sources. 
\section*{Acknowledgments}
We gratefully acknowledge financial support from the Deutsche Forschungsgemeinschaft (DFG, German Research Foundation) via projects MU4215/4-1 (CNLG), INST 95/1220-1 (MQCL) and INST 95/1654-1 (PQET), Germany’s Excellence Strategy (MCQST, EXC-2111, 390814868), the Bavarian Ministry of Economic Affairs (StMWi) via project 6GQT and the German Federal Ministry of Education and Research via the project 6G-life. F.S. thanks L. Hanschke, F. Lapatschek and O. Dincel for fruitful discussions.
\section*{Author contributions}
F.S., C.C., S.K.K., K.B., W.R., and F.B. set up and performed the experiments. E.Z.C. developed the theoretical model and performed the calculations. H.R. grew the sample and W.R. performed the fabrication and initial characterization. K.M. and J.J.F. provided expertise. F.S. and K.M. conceived the idea. All authors participated in the discussion and interpretation of the results. The manuscript was written by F.S. and E.Z.C. with contributions from all authors.

\section{Methods}
\subsection{Theoretical model}
\subsubsection{Resonance fluorescence}
We consider a two-level system, consisting of a ground state $\ket{0}$ and one excitonic state $\ket{1}$, whose energies are 0 and $\omega_\mathrm{X}$, respectively. Under resonant excitation,
with frequency $\omega_\mathrm{L} = \omega_\mathrm{X}$, the Hamiltonian in the Rotating Frame reads
\begin{equation}
    H_\mathrm{rf} = \Omega_\mathrm{X} (t) ( \sigma^\dagger + \sigma ) \,,
\end{equation}
where the transition operator is defined as $\sigma = \ket{0} \bra{1}$ and the driving amplitude has a Gaussian profile
\begin{equation}
\label{eq:gaussian-pulse}
   \Omega_\mathrm{X} (t) = \frac{\Theta}{\sqrt{2 \pi} \tau} \exp{- \frac{(t-t_0)^2}{2 \tau^2}} \,,
\end{equation}
where $\Theta$ and $\tau$ are the pulse area and length, respectively. The remaining
parameter $t_0$ corresponds to the time offset, that we set to $t_0 = 4 \tau$, which
enforces that the pulse excites the system after its initialization.

The dissipative processes are included in the system using a Lindblad-type Master Equation
\begin{equation}
    \partial_t \rho = -i[H_\mathrm{rf},\rho] + \frac{\gamma_\sigma}{2} \mathcal{L}_\sigma \rho \,,
\end{equation}
where $\mathcal{L}_c \rho = 2 c \rho \ud{c} - \ud{c} c \rho -   \rho \ud{c} c$ is the Lindblad term
corresponding to the operator $c$. In this case, $\gamma_\sigma$ corresponds to the decay rate of the excitonic state.
Now, we proceed to compute the frequency-filtered correlations. For such purpose, we can use
either the sensor method~\cite{delvalle2012} or the cascade formalism~\cite{gardiner1985,lopezcarreno2018}. 
We use the former for the computations.
In order to do so, we extend the Hilbert space by adding a sensor $\varsigma$ coupled
(with vanishing strength $\epsilon \rightarrow 0$) to the two-level system. The total Hamiltonian is then
\begin{equation}
    H = H_\mathrm{rf} + \Delta_1 \varsigma^\dagger \varsigma +
    \epsilon \big( \varsigma^\dagger \sigma + \mathrm{h.c.} \big),
\end{equation}
where $\Delta_1 = \omega_1 - \omega_\mathrm{L}$ is the central frequency of the sensor
referred to the laser frequency. The complete dynamics is then governed by the Master Equation
\begin{equation}
    \partial_t \rho = - i [H,\rho]  
    + \frac{\gamma_{\sigma}}{2}  \mathcal{L}_{\sigma} \rho +
     \frac{\Gamma}{2} \mathcal{L}_{\varsigma} \rho \,,
\end{equation}
where $\Gamma$ corresponds to the bandwidth of the Lorentzian filter. 
From the sensor observables, we obtain the frequency-resolved correlations, particularly the time-integrated population and two-photon correlations (see Supplemental Material for the details).
For Fig.~\ref{fig:theory}, we have chosen the filter frequency to be at resonance with the
two-level system transition: $\Delta_1 = 0$.
\subsubsection{Biexciton ladder}
The biexciton ladder consists of four levels: one ground state $\ket{\mathrm{cgs}}$, two excitonic states $\ket{\mathrm{X}_\mathrm{H}}$ and $\ket{\mathrm{X}_\mathrm{V}}$
(horizontal and vertical excitons, respectively), and
the two-exciton bound state $\ket{\mathrm{2X}}$. We shall use an index $i$ from 1 to 4,
which is assigned to each state as it was presented in the previous line.
We have four allowed transitions that are coupled to the electro-magnetic field: (i) the ones that couple
to the horizontally polarized ($\mathrm{H}$) light field ($\ket{\mathrm{2X}} \leftrightarrow \ket{\mathrm{X}_\mathrm{H}}$ and $\ket{\mathrm{X}_\mathrm{H}} \leftrightarrow \ket{\mathrm{cgs}}$),
and the ones that couple to the vertically polarized ($\mathrm{V}$) light field 
($\ket{\mathrm{2X}} \leftrightarrow \ket{\mathrm{X}_\mathrm{V}}$ and $\ket{\mathrm{X}_\mathrm{V}} \leftrightarrow \ket{\mathrm{cgs}}$).
From this we can easily see that the total contributions to the two polarizations are $X = \sigma_{21} + \sigma_{42}$ for $\mathrm{H}$ polarization and $Y = \sigma_{31} + \sigma_{43}$ for $\mathrm{V}$ polarization. In the presence of a coherent excitation (i.e., a laser)
with horizontal polarization (see the Supplemental Material for more details) and characteristic frequency $\omega_\mathrm{L}$,
the Hamiltonian (in the Rotating Frame) reads
\begin{multline}
H_\mathrm{X} = \Delta_\mathrm{X}  \sigma_{22} + \Delta_\mathrm{X}  \sigma_{33} + (2 \Delta_\mathrm{X} - E_b) \sigma_{44} + \\
\big[ \Omega_\mathrm{H} (t) (\sigma_{12} + \sigma_{24}) + \mathrm{h.c.} \big] \,,
\end{multline}
where the detuning is defined as $\Delta_\mathrm{X} = \omega_\mathrm{X} - \omega_\mathrm{L}$,
and $E_b$ is the biexciton binding energy.
In the two-photon excitation case, we set the laser frequency so that
the two photon transition matches the biexciton energy: $2 \omega_\mathrm{L} = 2
\omega_\mathrm{X} - E_b$ or, in turn, $\Delta_\mathrm{X} = -E_b/2$. 
For the excitation pulse, we choose again a Gaussian envelope, that is, the driving amplitude is $\Omega_\mathrm{H} (t) = 
\Omega_\mathrm{X} (t)$, as defined in Eq.~\eqref{eq:gaussian-pulse}.

We describe the dissipative dynamics using the Master Equation
\begin{equation}
	\label{eq:MEq}
	\dot{\rho} = -i[H_\mathrm{X},\rho] + \frac{\gamma_{\sigma}}{2} \big(  \mathcal{L}_{\sigma_{42}} \rho +
	\mathcal{L}_{\sigma_{21}} \rho + \mathcal{L}_{\sigma_{43}} \rho +
	\mathcal{L}_{\sigma_{31}} \rho 
	\big)
	\,,
\end{equation}
where we have assumed that all the transitions have the same decay rate $\gamma_{\sigma}$.
The detected polarization and the observed transitions are represented by the output field operator
\begin{align}
X_{\vec{\eta}} = \, & \vec{\eta} \cdot \vec{\sigma} =
    \begin{pmatrix}
        \eta_1 & \eta_2 & \eta_3 & \eta_4
    \end{pmatrix}
    \begin{pmatrix}
        \sigma_{21} \\
        \sigma_{42} \\
        \sigma_{31} \\
        \sigma_{43}
    \end{pmatrix}
\nonumber \\ 
= \, &
\eta_1 \, \sigma_{21} + \eta_2 \, \sigma_{42} + 
\eta_3 \, \sigma_{31} + \eta_4 \, \sigma_{43} \,,
\end{align}
where the elements of the observation vector $\vec{\eta}$ are, in general, complex numbers.

In order to compute the spectral correlations, we, again, extend the Hilbert space by adding a sensor $\varsigma$ coupled
(with vanishing strength $\epsilon \rightarrow 0$) to the output field
of our choice $X_{\vec{\eta}}$. The total Hamiltonian is then
\begin{equation}
    H = H_\mathrm{X} + \Delta_1 \varsigma^\dagger \varsigma +
    \epsilon \big( \varsigma^\dagger X_{\vec{\eta}} + \mathrm{h.c.} \big),
\end{equation}
where $\Delta_1 = \omega_1 - \omega_\mathrm{L}$ is the central frequency of the sensor
referred to the laser frequency. The complete dynamics is then governed by the Master Equation
\begin{multline}
    \partial_t \rho = - i [H,\rho] \\ 
    + \frac{\gamma_{\sigma}}{2} \big(  \mathcal{L}_{\sigma_{42}} \rho +
	\mathcal{L}_{\sigma_{21}} \rho + \mathcal{L}_{\sigma_{43}} \rho +
	\mathcal{L}_{\sigma_{31}} \rho 
	\big) +
     \frac{\Gamma}{2} \mathcal{L}_{\varsigma} \rho \,.
\end{multline}

For Fig.~\ref{fig:results}b, we fix the frequency of the sensor to $\Delta_1 = E_b /2$, the energy of the exciton-cgs transition,
and the observation vector to $\vec{\eta} = \begin{pmatrix} 0 & 0 & 1 & 0
\end{pmatrix}^\mathrm{T}$, that is, only the transition $\ket{\mathrm{X}_\mathrm{V}} \leftrightarrow \ket{\mathrm{cgs}}$ goes through the filter.
\subsection{Quantum dot sample}
The sample studied in this work consists of InGaAs quantum dots grown by molecular beam epitaxy using the Stranski-Krastanov growth method. To electrically control the charge state of the quantum dots, they are embedded in a p-i-n-i-n diode \cite{lobl2017}.
The carbon doped p-layer forms the front contact, while the back contact consists of a silicon doped n-layer.
The additional silicon doped n-layer between the quantum dot layer and the front contact reduces the built-in electric field and thus enables photoluminescence at low applied voltages. To enhance collection efficiency, a distributed Bragg reflector consisting of 17 pairs of alternating AlAs and GaAs was grown below the diode structure, which in conjunction with the sample surface forms a weak planar $\lambda/2$ cavity. The quantum dot layer is centered in the field maximum of this cavity. For all experiments, the sample is cooled to $\SI{1.7}{\kelvin}$ in a closed-cycle cryostat. The exciton transition of the studied quantum dot has an energy of $E_{\mathrm{X}_\mathrm{V}}= \SI{1354.1}{\milli \eV}$, from where the biexciton is red-detuned by its binding energy $E_b= \SI{2.1}{\milli \eV}$.

\subsection{Pulse generation}
The picosecond laser pulses used for driving the cgs-biexciton transition are derived from a femtosecond pulsed Ti:sapph laser using two folded 4f pulse shapers \cite{weiner1988} in a serial configuration.
Both pulse shapers consist of a plane ruled diffraction grating with 1800 grooves/mm and a lens with $\SI{300}{\milli \meter}$ focal length.
In the Fourier plane of the pulse shapers an adjustable mechanical slit mounted on a translation stage and a mirror are positioned.
This setup allows for precise frequency control of the pulse's center frequency via the position of the slit, 
and for control of the pulse duration via the slit width.
The pulse durations are measured with an intensity autocorrelator. 
By combining the two pulse shapers in series we can further suppress stray light originating from diffuse stray-processes on the surface of the slit, improving the maximal suppression of the laser in the experiment.

\subsection{Polarization and frequency filtering}
To achieve the high signal-to-noise ratio necessary to measure such low $g^{(2)} \left[ 0; \Gamma_\nu \right]$ values in the experiment, we employ multiple techniques of spectral and polarization filtering.
In all measurements, we use cross-polarized filtering to reduce the laser intensity in the detection path.
To this end, we insert two orthogonal thin-film linear polarizers in the excitation and detection path of the micro-photoluminescence setup. To correct for birefringence of the optics and the sample, we additionally have a quarter-wave plate positioned in the detection path, in front of the linear polarizer.
To spectrally filter the emission, we employ two different types of grating-based filters.
The first filter is based on the pulse shaper design used for excitation, here in an unfolded design consisting of two gratings with 1800 grooves/mm and two lenses, each with $\SI{300}{\milli \meter}$ focal length.
Again, an adjustable slit is positioned in the Fourier plane on a translation stage, where the position determines the center transmission frequency and the slit width determines the bandwidth of the filter.
The bandwidth $\Gamma_\nu$ for different slit widths is determined by measuring the transmission with a tunable continuous-wave laser and fitting it with a super-Gaussian function \cite{decker1995}.
This filter enables transmission bandwidths between $\SI{281}{\giga \hertz}$ and $\SI{41}{\giga \hertz}$. 
For narrower transmission bandwidths, we employ a second filter design, where the collimated light is sent onto a transmission grating with 1500 grooves/mm. Filtering is achieved by coupling the diffracted light of the desired frequency into a single-mode fiber, where the fiber core acts as the spatial filter of the spatially separated frequencies. This filter has a bandwidth of $\SI{23}{\giga \hertz}$. 
To further narrow the bandwidth, an etalon with a free-spectral range of $\SI{60}{\giga \hertz}$ was inserted into the optical path of the transmission grating filter. 
This reduces the filter's bandwidth to $\SI{1.4}{\giga \hertz}$ with a Lorentzian shape.\\

\subsection{Time-resolved measurements}
The time-resolved measurements are taken by sending the filtered quantum dot emission via a Hanbury Brown and Twiss setup to  two superconducting nanowire single photon detectors. The coincidences are correlated using a streaming time-to-digital converter. For the correlation measurements presented in the manuscript, we display coincidences within $\SI{5}{\pico \second}$ wide bins. 
We determine the degree of second-order coherence at time delay $t=0$ by dividing the summed counts of the center peak by the average counts of the two neighboring side peaks at $t=\pm \SI{13.1}{\nano \second}$.
To exclude the peaks originating from the reflected signal, we sum all peaks over a total time window of $\SI{6.5}{\nano \second}$ centered at their respective positions.
This corresponds to more than 22 times the radiative lifetime of the exciton transition and includes more than 99.98 \% of the total emission.
The errors for the $g^{(2)} \left[ 0; \Gamma_\nu \right]$ values given in the manuscript represent the $1\sigma$ confidence interval. We determine the uncertainty by assuming Poissonian counting statistics for the summed events of the center and neighboring peaks and applying Gaussian error propagation.

\bibliography{bibliography}

\appendix 
\onecolumngrid
\newpage

\section{Details of the theoretical model}
\label{app:appendix1}
\subsection{Time-integrated photon correlations}
We are particularly interested in the time-integrated two-photon correlations
\begin{equation}
\label{eq:filtered-g2}
g^{(2)} [0;\Gamma] = \frac{\int_0^T \int_0^T \mathcal{G}_\Gamma^{(2)} (t_1, t_2) \, dt_1 dt_2}{
(\int_0^T n_\Gamma (t) \, dt )^2} \,,
\end{equation}
where $T$, the integration time, is taken to infinity ($T \rightarrow \infty$). 
The frequency-filtered population and the second-order Glauber correlation function
are defined in terms of the sensor correlations (see Methods) as 
\begin{equation}
    n_\Gamma (t) = \Big(\frac{\Gamma}{2 \epsilon} \Big)^2 \av{\varsigma^\dagger (t) \varsigma (t) }
\,,
\end{equation}
and
\begin{equation}
 \mathcal{G}_\Gamma^{(2)} (t_1,t_2) = \Big(\frac{\Gamma}{2 \epsilon} \Big)^4 \av{\mathcal{T}_{-}
 [\varsigma^\dagger (t_1) \varsigma^\dagger (t_2)] \mathcal{T}_+[ \varsigma (t_2) \varsigma (t_1)]} \,,
\end{equation}
where $\mathcal{T}_{\pm}$ chronologically orders the operators from right to left,
and left to right, respectively. The numerator and denominator of Eq.~\eqref{eq:filtered-g2},
that is, the time-integrated two-photon correlation and population, respectively,
can be efficiently computed using the methods developed in Ref.~\cite{bermudezfeijoo2025}.
When the filter is large enough, that is, $\Gamma \rightarrow \infty$, we recover the unfiltered
photon statistics. In particular, for the two-level system emission, the correlation function converges to
\begin{equation}
  \mathcal{G}_\sigma^{(2)} (t_1,t_2) =\mathcal{G}_{\Gamma \rightarrow \infty}^{(2)} (t_1,t_2) =
 \av{\mathcal{T}_{-}
 [\sigma^\dagger (t_1) \sigma^\dagger (t_2)] \mathcal{T}_+[ \sigma (t_2) \sigma (t_1)]} \,, 
\end{equation}
which corresponds to the result shown in Fig.~1a.
\subsection{Biexciton ladder polarization}
As explained in the Methods section, the biexciton ladder is coupled to the two linear
polarizations, that is, $X = \sigma_{21} + \sigma_{42}$ couples to the $\mathrm{H}$ polarization, whereas $Y = \sigma_{31} + \sigma_{43}$ couples to the $\mathrm{V}$ polarization. 
The input laser field can have any arbitrary polarization, characterized by the
vector $\hat{u}_\mathrm{in}$
\begin{equation}
    \hat{u}_\mathrm{in} =
    \begin{pmatrix}
        \cos{\theta} \\ \sin{\theta} \, e^{i \phi} 
    \end{pmatrix}
    \,,
\end{equation}
which includes all the polarization states we will consider (fully polarized light), such as
linearly polarization $\hat{u}_{\theta}$ (with $0 \leq \theta < \pi$ $\phi = 0$) and
circularly (left and right) polarized light $\hat{u}_\mathrm{L,R}$ (with $\theta = \pi/4$ and $\phi = \pm \pi /2 $, respectively).
As we previously discussed, each branch of the exciton ladder couples to the horizontal and
vertical components of the light field, then the effective amplitude of each branch is 
$\Omega_\mathrm{H,V} (t) = \Omega_0 (t) (\hat{u}_\mathrm{in}^* \cdot \hat{v}_\mathrm{H,V})$,
where $\hat{v}_\mathrm{H} = \hat{u}_0$ and $\hat{v}_\mathrm{V} = \hat{u}_{\pi/2}$, which is obtained
by projecting the input polarization vector onto the horizontal and vertical vectors, and
$\Omega_0 (t)$ is the effective laser amplitude. For all the computation, we have assumed the laser polarization to be horizontal. 
\section{Lifetime}
\begin{figure}
    \centering
    \includegraphics{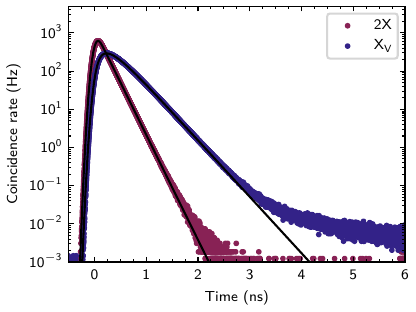}
    \caption{\textbf{Lifetime measurements.} Time-resolved biexciton and exciton photoluminescence upon excitation of the biexciton with a $\pi$-pulse. The exciton state population slowly builds up as the biexciton decays. Fitting the two curves (black lines), we extract $\tau_\mathrm{2X} = \SI{158}{\pico \second}$ and $\tau_\mathrm{X_V}= \SI{294}{\pico \second}$. }
    \label{Sfig:lifetime}
\end{figure}

We measure the lifetimes of the two transitions of the biexciton-exciton cascaded decay upon excitation of the biexciton via the two-photon transition with a pulse of area $\pi$ and length \SI{5}{\pico \second}. A photodiode signal derived from the laser pulse is correlated with either the biexciton (wine dots) or exciton (indigo dots) emission as shown in \cref{Sfig:lifetime}.
To filter the emission, we use a spectral filter with a width $\Gamma_\nu= \SI{98}{\giga \hertz}$.
We fit the curves with a convolution of the solution to their coupled rate equations and a Gaussian function (black lines). 
Here, the width of the Gaussian which corresponds to the timing resolution of our setup is left as a free fit parameter. 
For the biexciton emission this yields the expected monoexponential decay with a lifetime $\tau_\mathrm{2X}= \SI{158}{\pico \second}$. 
For the exciton we observe the expected buildup and subsequent decay and extract a lifetime of $\tau_\mathrm{X_V}= \SI{294}{\pico \second}$. In addition, the measurement exhibits a very weak second decay, which which we attribute to a very low but finite probability of a spin flip and thus creation of a dark exciton \cite{dalgarno2005}.
\section{Signal-to-noise ratio}
\begin{figure}
    \centering
    \includegraphics{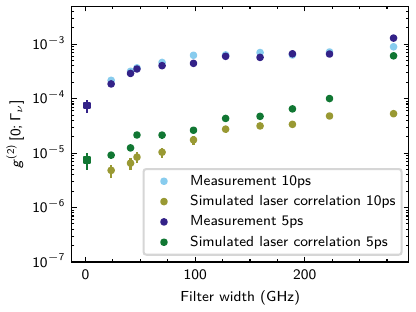}
    \caption{\textbf{Signal-to-noise ratio estimation.} Measured data of the filter dependent $g^{(2)}\left[0; \Gamma_\nu \right]$ from \cref{fig:results}a from the main manuscript (indigo and cyan). The green and olive data points represent our lower bound estimate of the $g^{(2)}\left[0; \Gamma_\nu \right]$, which is derived from the simulated correlations of the laser leakage and a perfect single-photon source.   }
    \label{Sfig:SNR}
\end{figure}

To be able to resolve the instantaneously emitted photon in the second-order correlations, it is crucial to maximize the signal-to-noise ratio of the experiment.
To optimize this ratio, we chose a specific quantum dot transition (the exciton emission in a cascaded four-level system) and apply multiple polarization and spectral filtering techniques, as described in the manuscript.
To ensure that we are measuring the instantaneous photon instead of undesired correlation events from signal and noise, we investigated the origin of the residual noise in our setup and identified it to be dominated by laser leakage into the emission signal.
Therefore, right before each correlation measurement, we took noise count rate measurements with the quantum dot turned out of the photoluminescence regime by biasing the sample diode, but with the excitation pulse turned on.
The measured count rate includes both the residual laser leakage counts and the intrinsic dark count rate of the detectors, which is in our case, due to the use of low noise superconducting nanowire detectors, lower than the laser leakage count rate.
For the measurements in the manuscript, the typical ratios between noise counts and quantum dot counts lie between 1:20000 and 1:400000. 
Only for the measurement using the widest filter and the shorter pulse length, significant amounts of laser light start to leak in, resulting in deterioration of the ratio to 1:3400.
To estimate the minimum obtainable $g^{(2)}\left[0; \Gamma_\nu \right]$, we performed a Monte-Carlo simulation, where we assumed a perfect single-photon source without any two-photon events.
We use the measured quantum dot count rates and an additional Poissonian noise source with the measured laser leakage count rate.
We determine the resulting $g^{(2)}\left[0; \Gamma_\nu \right]$ by simulating the autocorrelation of the photon stream in a Hanbury Brown Twiss experiment, which is displayed by the green and olive data points for pulse durations of \SI{5}{\pico \second} and \SI{10}{\pico \second} in \cref{Sfig:SNR}, respectively.
For all data points but one, we observe that our estimate of the lowest possible $g^{(2)}\left[0; \Gamma_\nu \right]$ with the measured noise of our experimental setup is significantly lower than the experimentally measured values.
This means that the second-order coherence measurements presented in the manuscript are dominated by the intrinsic multi-photon error rate of the system and not by coincidences with dark or laser leakage counts.
Only for the short pulse length and the widest filter we observe that the lower bound estimate is of a similar order as the measured $g^{(2)}\left[0; \Gamma_\nu \right]$, which negatively affects the measured value. 

\section{Blinking}
\begin{figure}
    \centering
    \includegraphics{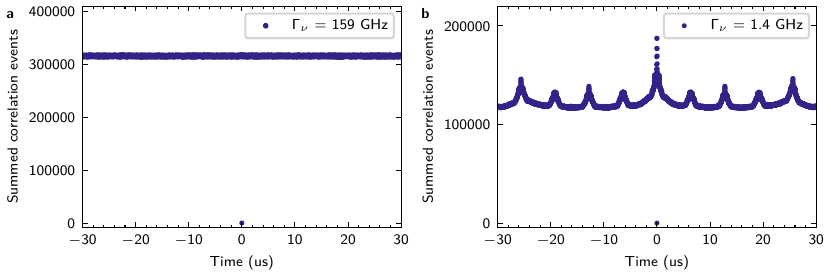}
    \caption{\textbf{Blinking analysis.} \textbf{a}, Summed counts of the individual peaks of the autocorrelation measurement with the filter width $\Gamma_\nu = \SI{159}{\giga \hertz}$ are constant for all side peaks. Only the center peak at zero time delay differs. \textbf{b}, Summed counts of the individual peaks for the filter width $\Gamma_\nu = \SI{1.4}{\giga \hertz}$ show a periodic blinking on a microsecond timescale.}
    \label{Sfig:Blinking}
\end{figure}

As discussed in the manuscript, the studied sample is subject to periodic electric noise which results in blinking for the narrowest filter as displayed in \cref{Sfig:Blinking}.
We investigate this by summing the counts of all individual peaks of the correlation measurement.
For the wider filter of $\Gamma_\nu = \SI{159}{\giga \hertz}$ this exhibits a flat line, as seen in \cref{Sfig:Blinking}a.
This means that we observe no blinking in this measurement configuration.
In contrast, for the very narrow etalon filter measurement in the manuscript, we observe a very clear and periodic blinking in the second-order intensity autocorrelation measurement on microsecond timescales as seen in \cref{Sfig:Blinking}b.
Because of the periodicity and multiple discrete frequencies in the MHz range, we attribute this blinking to periodic electric noise affecting the quantum dot. 
Together with the quantum confined Stark effect \cite{warburton2000a}, this leads to spectral diffusion on MHz timescales.
When the emission is narrowly spectrally filtered, this becomes visible as the observed periodic bunching.
The electric noise acting on the quantum dot can be minimized by improved wire shielding, using electric lowpass filters, and lower noise DC sources to bias the diode structure.
If the residual electric noise is minimized until the emission linewidth becomes Fourier-limited, this also will inhibit spectral diffusion and result in the restoration of the constant summed correlation events for each individual peak on long timescales.
This would remove the decrease of the pulse-wise summed counts of the neighboring peaks for increasing time delay, potentially allowing for the measurement of even lower $g^{(2)}\left[0; \Gamma_\nu \right]$ values.

\end{document}